\documentclass[twocolumn,twoside,slac_two]{revtex4}
\usepackage{graphicx}
\usepackage{fancyhdr}
\begin{document}

\title{Radiative Corrections in K, D, and B decays}


\author{Vincenzo Cirigliano}
\affiliation{California Institute of Technology, Pasadena, CA 91125, USA   {\it and}  \\ 
Theoretical Division, Los Alamos National Laboratory,  Los Alamos, NM 87545,  USA}

\begin{abstract}

In this talk I  review recent developments in the calculation of 
radiative corrections to $K, D, B$  meson  decays and discuss 
their impact on precision phenomenology. 

\end{abstract}

\maketitle

\thispagestyle{fancy}

\section{Motivation}

The calculation of electromagnetic radiative corrections (RC) to a given observable 
is typically needed when the corresponding experimental measurement 
reaches the precision of a few percent.   
In the context of flavor physics, 
RC are truly at the basis for meaningful $\%$-level phenomenology for two-fold reasons:

\begin{itemize}

\item One one hand,  RC  are essential in extracting  fundamental Standard Model parameters 
(like the CKM matrix elements) or weak hadronic matrix elements 
from precisely measured decay rates. 

\item On the other hand, recent experience shows that knowledge of 
radiative effects (such as emission of real photons) and their proper simulation 
are essential in performing  {\em precision measurements}  of branching fractions.  
In fact, real photon emission distorts the spectra of  "primary" 
decay products and thus affects  the distribution of
kinematic quantities used to select and count signal
events~\cite{RadCorr,CGatti}. The inclusion of RC in the experimental 
fits  is  known to produce shifts in the measured fractions at the several percent level. 

\end{itemize}

The relevance of RC has been widely  recognized in the K physics community,    
where high statistics experiments have triggered theoretical studies on the subject 
throughout the past decade.  
In  heavy flavor physics this issue has received comparatively less attention. 
However,  experiments at B and charm factories are 
approaching (in some channels) an accuracy level that 
requires a satisfactory understanding of radiative effects.    

In this talk I review the theoretical framework for the study of RC to meson decays 
and describe recent theoretical efforts  to obtain  model-independent results 
despite the complications due to strong interactions.

\section{Radiative Corrections in Meson Decays: Overview}

 \subsection{Theoretical Framework}

Within the Standard Model (and its extensions)  weak decays of mesons 
are studied via an effective lagrangian which factorizes the short- and long-distance physics:
\begin{equation}
{\cal L}_{eff} =   G_F \, \sum_{n}  \  C_n   \,  O_n  \ . 
\label{eq:Leff}
\end{equation}
The Wilson coefficients $C_n$ encode information on short-distance degrees 
of freedom ($W$,$Z$,$t$  masses, for example), while the local dimension-six operators $O_n$ are 
constructed  out  of  "light" fields (quarks, leptons, gluons, photons).   

In this framework, radiative corrections are accounted for by keeping track of 
$O(\alpha)$ effects in the calculation of both Wilson coefficients  and hadronic 
matrix elements  $\langle f|  O_n | H \rangle$, eventually resulting in  $O(\alpha)$ 
corrections to the decay rate  $\Gamma (H\to f)$.  The effective lagrangian framework 
leads to a natural  separation into short- and long-distance corrections.  \\
(i) {\em Short distance} (SD) corrections to the Wilson 
coefficients are calculated in perturbation theory and 
are  by now  well known for both semi-leptonic~\cite{SDcorr1}  and 
non-leptonic~\cite{SDcorr2}  operators.
They typically result in large UV logs  $\delta C_n \sim \alpha/\pi  \log \left( M_Z/M_H \right)$, 
implying  $\delta \Gamma_{\rm SD}/ \Gamma \sim \, {\rm few} \, \% $. \\
(ii) {\em Long distance} (LD) corrections in general cannot be calculated in perturbation theory.
The matrix elements $\langle f|  O_n | H \rangle$  pose great challenges  already 
without inclusion of $O(\alpha)$ effects, i.e. without virtual photons.    
Rigorous results can be obtained only in certain ranges of external momenta, corresponding to 
various hadronic Effective Field Theories (EFT), for example 
 Chiral Perturbation Theory (ChPT), Heavy 
Quark Effective Theory (HQET), Heavy Hadron ChPT (HHChPT).
In these cases it is possible to at least formally match from the quark-level to the 
hadronic EFT, thus generating  $O(\alpha)$ effective hadronic vertices  
without introduction of large logarithms. 

Despite the hadronic uncertainties, an important class of long distance 
corrections, the infrared (IR) effects,  can be treated in a model-independent way.
IR photons are defined as those with  a wavelength  much larger than the typical size of hadrons  
and therefore are  not sensitive to the hadronic structure. 
Their  effects  can be calculated in the approximation of point-like hadrons, using 
only gauge invariance as the guiding principle in  constructing the interaction lagrangian.
The calculation of virtual photon corrections in the point-like approximation leads (as in 
standard QED)  to IR divergent matrix elements and decay rates. 
One obtains IR-finite results only for the  observable {\em photon-inclusive}  decay rates such as 
\begin{equation}
d \Gamma^{\rm incl} (E_\gamma^{\rm cut}) = d \Gamma_{H \to f}  + d \Gamma_{H \to f + \gamma} +
 d \Gamma _{H \to  f + 2 \gamma} + \dots   
\end{equation} 
with $  \sum_{n} E_\gamma^{(n)}  <  E_\gamma^{\rm cut}  $. 
The typical size of long-distance corrections is then given by 
 $\delta \Gamma_{\rm LD}/ \Gamma \sim \alpha/\pi \log  \left( M_H/E_\gamma^{\rm cut}  \right)$. 
When $E_\gamma^{\rm cut}  \ll M_H$ the IR corrections are quite large and one needs 
to sum the effect to all orders in perturbation theory~\cite{Weinberg:1965nx,Yennie:1961ad}.

\subsection{General parameterization} 

Denoting with $\xi$ the collection of independent variables characterizing the non-radiative 
$H \to f$ decay and with $E_\gamma^{\rm cut} (\xi)$  the  photon energy cut  (in general $\xi$-dependent),  the IR safe decay distribution can be written as follows
\begin{eqnarray}
\frac{d \Gamma^{\rm incl}}{d \xi}  &=& \left( \frac{\bar{E}}{M_H} \right)^{b(\xi)}   \  
\frac{d \Gamma^{(0)}}{d \xi}  \, (1 + \delta_{SD}) \nonumber \\
&+& \delta_{LD} \left(\xi, E_\gamma^{\rm cut} (\xi)  / \bar{E} \right)  \ , 
\label{eq:general}
\end{eqnarray} 
where 
\begin{equation}
b = - \frac{1}{8 \pi^2}  \sum_{m,n} \eta_m \eta_n e_m e_n  \,  \beta_{mn}^{-1}  \ 
\log \frac{1 + \beta_{mn}}{1 - \beta_{mn}}   ~ . 
\label{eq:b}
\end{equation}
The sum in Eq.~(\ref{eq:b}) runs over pairs of charged particles appearing in the  
initial and final states, $e_{m,n}$ are their electric charges, while  $\eta_m=+1$ 
for out-going particles and $\eta_m=-1$ for in-going particles. 
$\beta_{mn}$ denotes the relative velocity  of particles $m$ and $n$ in the rest 
frame of either of them~\cite{Weinberg:1965nx}: 
\begin{equation}
\beta_{mn} =  \left[ 1 -  \frac{m_n^2 m_m^2}{(p_n \cdot p_n)^2} \right]^{1/2}  \ . 
\end{equation}
Eq.~(\ref{eq:general}) requires some comments:
\begin{itemize}

\item   $\bar{E} \ll M_H$ is an arbitrary soft scale, used to define IR photons
\footnote{Ultimately, the inclusive rate does not depend on $\bar{E}$: 
the  $\bar{E}$ dependence cancels 
between the IR factor $(\bar{E}/M_H)^b$  and $\delta_{LD}$. }.
IR effects from real and virtual photons can be re-summed to all orders in 
perturbation theory~\cite{Weinberg:1965nx}  and result  in the overall factor 
$\left( \frac{\bar{E}}{M_H} \right)^{b(\xi)}$. This effect is model-independent 
and for $E_\gamma^{\rm cut} (\xi) \sim \bar{E} \ll M_H$  provides the largest 
long-distance correction. 

\item  $\delta_{SD}$ is the model-independent short distance correction and   
always appears  as a multiplicative correction to the "Born" rate   
$\frac{d \Gamma^{(0)}}{d \xi} $.

\item    $\delta_{LD} \left(\xi, E_\gamma^{\rm cut} (\xi)  / \bar{E} \right)  $ is the long-distance, structure dependent correction. 
In general it does not factorize in the product of Born rate times a correction. 
In experimental setups where 
$E_\gamma^{\rm cut} (\xi) \sim E_\gamma^{\rm MAX} (\xi) \sim O( M_H)$ , 
this is the dominant LD correction. 

\end{itemize}

A  parameterization similar to Eq.~(\ref{eq:general})  is possible for the photon-energy distribution:
\begin{equation}
\frac{d \Gamma^{\rm incl}}{dE_\gamma \, d \xi} = \left( \frac{E_\gamma}{M_H} \right)^{b(\xi)}   \, 
\frac{b(\xi)}{E_\gamma}  
\frac{d \Gamma^{(0)}}{d \xi}  +
\tilde{\delta}_{LD} \left(\xi, E_\gamma \right)  \ . 
\label{eq:general2}
\end{equation}
The first term describes the universal soft-photon spectrum and is fixed by 
gauge invariance and kinematics. It is crucial in describing the soft radiative tails in 
experimental Monte Carlo studies~\cite{CGatti}. 
The second term starts at  $O(E_\gamma^0)$  and  describes the hard-photon spectrum.  
The impact of  $\tilde{\delta}_{LD} \left(\xi, E_\gamma \right)$ in simulations 
depends on the specific experimental setup. 

In the parameterization of Eqs.~(\ref{eq:general}) and (\ref{eq:general2}) 
$\delta_{LD}$ and $\tilde{\delta}_{LD}$ are 
the  quantities  sensitive to the hadronic structure, and therefore the hardest to calculate. 
In Kaon decays, ChPT provides a universal tool to  treat both virtual and real photon 
effects over all of the available phase space.      
The framework to treat both semileptonic ~\cite{Knecht:1999ag}
and non-leptonic~\cite{Ecker:2000zr} decays has been fully developed and 
explicit calculation exist for $K \to \pi \pi$~\cite{k2pi}, 
$K \to \pi \pi \pi $~\cite{k3pi}, 
$K \to  \pi \ell \nu $~\cite{radcorr1,radcorr1bis},  
$K \to \pi \pi \ell \nu $~\cite{kl4}. 
On the other hand, in most  B (D) decays, rigorous  EFT treatments only work in 
corners of phase space.  
In what follows I describe a few examples of  calculations 
of  radiative corrections  to exclusive K, D, and B decays and their impact on phenomenology.

\section{Recent progress in exclusive modes}

\subsection{  $K \to \pi \ell  \bar{\nu}_\ell  (\gamma)$ and $V_{us}$ }

\begin{table*}[t]
\begin{center}
\begin{tabular}{|r||r||r|r|}
\hline
&  $\Delta^K_{SU(2)}$ (\%) 
& \multicolumn{2}{|c|}{
 $\Delta^{K \ell}_{\rm EM}(\%)$}     \\
&  &
\multicolumn{2}{|c|}{ 3-body $\qquad\qquad$ full}  \\
\hline
$K^{+}_{e3}$ & 2.31 $\pm$ 0.22 ~ \cite{Gasser:1984ux,radcorr1}
&  -0.35  $\pm$ 0.16 ~ \cite{radcorr1}
& -0.10 $\pm$ 0.16 ~ \cite{radcorr1} \\
\hline 
%             &              &                    &                    \\
$K^{0}_{e 3}$ &  0     &   +0.30  $\pm$  0.10 ~ \cite{radcorr1bis}
&   +0.55 $\pm$ 0.10   ~ \cite{radcorr1bis}  \\
              &             &                      & 
+0.65 $\pm$ 0.15  ~ \cite{radcorr2} \\
\hline 
$K^{+}_{\mu 3}$ & 2.31 $\pm$ 0.22 ~ \cite{Gasser:1984ux,radcorr1}
 &                    &              \\
\hline 
%                &                 &               &                   \\
$K^{0}_{\mu 3}$ &  0               &               & 
+0.95 $\pm$ 0.15  ~ \cite{radcorr2} \\
\hline
\end{tabular}
\caption{
Summary of SU(2) and radiative correction factors for various $K_{\ell 3}$ 
decay modes.  Refs.~\cite{Gasser:1984ux,radcorr1,radcorr1bis} work within 
chiral perturbation theory to order $p^4, e^2 p^2$, while 
Ref.~\cite{radcorr2} works within a hadronic model for Kaon electromagnetic 
interactions.  
}
\label{tab:radcorr}
\end{center}
\end{table*}

In this section I briefly review the extraction of $V_{us}$ from $K_{\ell 3}$ decays 
and the impact of radiative corrections.    
The decay rates for all  $K_{\ell 3}$ modes ($K={K^\pm, K^0}, 
\ell=\mu,e$)  can be written 
compactly as follows:
\begin{eqnarray}
\Gamma (K_{\ell 3 [\gamma] })  &= & \frac{G_F^2 \,  S_{\rm ew} 
\,  M_K^5}{128 \pi^3} 
\, C^K  I^{ K \ell} (  \lambda_i )  \times   \nonumber \\
| V_{us}   &\times & f_{+}^{K^0 \pi^-} (0)|^2  
\times \Big[ 
1 + 2\, \Delta^K_{SU(2)} + 2\,  \Delta^{K \ell}_{\rm EM} 
\Big] \ .  \nonumber \\
\label{eq:masterkl3}
\end{eqnarray}
Here $G_F$ is the Fermi constant as extracted from muon decay,
$S_{\rm ew} = 1 + \frac{2 \alpha}{\pi} \left( 1 -\frac{\alpha_s}{4 \pi} \right)\times 
\log \frac{M_Z}{M_\rho}  + O (\frac{\alpha \alpha_s}{\pi^2})$ 
represents the short distance electroweak correction to semileptonic
charged-current processes~\cite{SDcorr1},  $C^K$ is a Clebsh-Gordan coefficient equal
to 1 (1/$\sqrt{2}$) for neutral (charged) kaon decay, while $I^{ K \ell} (
\lambda_i )$ is a phase-space integral depending on slope and
curvature of the form factors. The latter are 
defined by the QCD matrix elements
\begin{eqnarray}
&\langle \pi^j (p_\pi) | \bar{s} \gamma_\mu u | K^i (p_K) \rangle = & \nonumber \\
&f_{+}^{K^i \pi^j} (t)  \, (p_K + p_\pi)_\mu  + 
f_{-}^{K^i \pi^j} (t)  \, (p_K - p_\pi)_\mu ~. &
\end{eqnarray}
As shown explicitly in Eq.~(\ref{eq:masterkl3}), it is convenient 
to normalize the form factors of all channels to $f_{+}^{K^0 \pi^-} (0)$, 
which in the following will simply be denoted by $f_{+}(0)$.
The channel-dependent terms 
$\Delta^K_{SU(2)}$ and $\Delta^{K \ell}_{\rm EM}$ represent
the isospin-breaking and long-distance electromagnetic corrections,
respectively. A determination of $V_{us}$ from $K_{\ell 3}$ decays at
the $1\%$ level requires $\sim 1 \%$ theoretical control 
on  $ f_{+}(0)$ as well as the inclusion of 
$\Delta^K_{SU(2)}$ and $ \Delta^{K \ell}_{\rm EM}$. 

The natural framework to analyze these corrections is provided by
chiral perturbation
theory~\cite{Weinberg:1978kz,Gasser:1983yg,Gasser:1984gg} (CHPT), the
low energy effective theory of QCD.  Physical amplitudes are
systematically expanded in powers of external momenta of
pseudo-Goldstone bosons ($\pi, K , \eta$) and quark masses.  When
including electromagnetic corrections, the power counting is in
$(e^2)^{m} \, (p^2/\Lambda_\chi^2)^{n}$, with $\Lambda_\chi \sim 4 \pi
F_\pi$ and $p^2 \sim O(p_{\rm ext}^2 , M_{K,\pi}^2) \sim O(m_q) $.  To
a given order in the above expansion, the effective theory contains a
number of low energy couplings (LECs) unconstrained by symmetry alone.
In lowest orders one retains model-independent predictive power, as
these couplings can be determined by fitting a subset of available
observables. Even in higher orders the effective theory framework
proves very useful, as it allows one to bound unknown or neglected 
terms via power counting and dimensional analysis arguments.

Strong isospin breaking effects $O(m_u - m_d)$ were first studied to
$O(p^4)$ in Ref.~\cite{Gasser:1984ux}.  Both loop and LECs contributions
appear to this order. Using updated input on quark masses and the
relevant LECs, the results quoted in Table~\ref{tab:radcorr} for
$\Delta^K_{SU(2)}$ were obtained in Ref.~\cite{radcorr1}.

Long distance electromagnetic corrections were studied within CHPT to
order $e^2 p^2$ in Refs.~\cite{radcorr1,radcorr1bis}.  To this order,
both virtual and real photon corrections contribute to $\Delta^{K
\ell}_{\rm EM}$.  The virtual photon corrections involve (known) loops
and tree level diagrams with insertion of $O(e^2 p^2)$ LECs.  Some of
The relevant LECs have been estimated in~\cite{moussallam} and more recently in 
\cite{descotes} using large-$N_C$
techniques. These works show that once the large UV logs have been isolated, 
the residual values of the LECs are well within the bounds implied by 
 naive dimensional analysis.     The resulting matching  uncertainty is
reported in Table~\ref{tab:radcorr}, and does not affect the
extraction of $V_{us}$ at an appreciable level.

Radiation of real photons is also an important ingredient in the
calculation of $\Delta^{K \ell}_{\rm EM}$, because only the inclusive
sum of $K_{\ell3}$ and $K_{\ell 3 \gamma}$ rates is infrared finite to
any order in $\alpha$. Moreover, the correction factor depends on the
precise definition of inclusive rate.  In Table~\ref{tab:radcorr} we
collect results for the fully inclusive rate (``full'') and for the
``3-body'' prescription, where only radiative events consistent with three-body
kinematics are kept. CHPT power counting implies that to order
$e^2 p^2$ one has to treat $K$ and $\pi$ as point-like (and with
constant weak form factors) in the calculation of the radiative rate,
while structure dependent effects enter to higher order in the
chiral expansion ~\cite{kl3rad}.

Radiative corrections to $K_{\ell 3}$ decays
have been recently calculated also outside the CHPT
framework~\cite{radcorr2,radcorr3}.  Within these schemes, the UV
divergences of loops are regulated with a cutoff (chosen to be around
1 GeV).  In addition, the treatment of radiative decays includes part
of the structure dependent effects, introduced by the use of form
factors in the weak vertices.  Table~\ref{tab:radcorr} shows that
numerically the ``model'' approach of Ref.~\cite{radcorr2} agrees
rather well with the effective theory.

\begin{figure}[h]
\centering
\includegraphics[width=80mm]{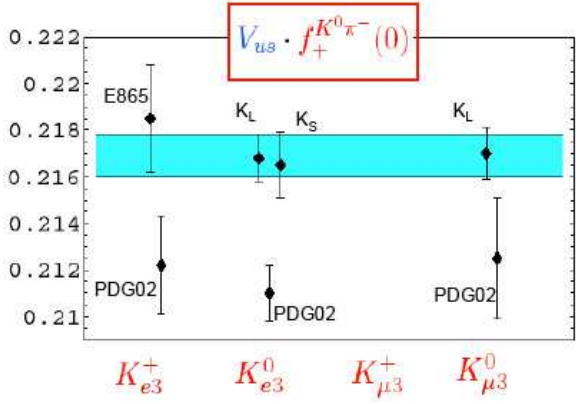}
\caption{The quantity $V_{us} \times f_+(0)$ as extracted from various 
$K_{\ell 3}$ decay modes.  The theoretical input needed to 
obtain these plots is discussed in the text. All the experimental input is taken 
from~\cite{blucher-marciano}. The shaded horizontal band represent  the average 
over all modes (including correlations) with one-sigma error.  
 For comparison, results based on "old" (PDG02)  $K_{\ell 3}$ 
measurements are also shown.} \label{fig:fvus}
\end{figure}

Finally, it is worth stressing that the 
consistency of the calculated isospin-breaking and electromagnetic 
corrections can be tested by experimental data by comparing the determination of
$V_{us} \times f_{+} (0)$ from various decay modes,
as shown in Fig.~\ref{fig:fvus}.
The average is~\cite{blucher-marciano}
\begin{equation}
V_{us}  \times  f_+(0)  =  0.2169  \pm  0.0009 ~ . 
\end{equation}
An analysis  of the uncertainties reveals that 
$\delta V_{us}/V_{us} = 0.2 \% (\Gamma)  \pm 0.35 \%  (I^{K\ell})  \pm  0.1 \% (\Delta_{EM}^{K\ell})$  
confirming that the  electromagnetic corrections are at the moment well under control. 
In the final  extraction of $V_{us}$ the dominant uncertainty ($\sim 1 \%$) comes 
form the hadronic form factor $f_+(0)$~\cite{blucher-marciano}.

%%%%%%%%%%%%%%%%%%%%%%%%%%%%%%%%%%%%%%%%%%%%%%%%%%%%%%%%%%%%%%%%%%%%%%%%%%%%%

\subsection{ $B (D) \to  P_1 P_2$}
 
Radiative corrections in heavy meson decays pose a harder theoretical problem, 
due to the lack of a universally valid effective theory (as it is ChPT in Kaon decays). 
Nonetheless, as remarked earlier,  it is certainly possible to calculate 
the important IR effects using the approximation of point-like hadrons. 
A first step in this direction has been made in Ref.~\cite{Baracchini:2005wp}, 
that studied the non-leptonic decays $B(D) \to P_1 P_2$ of heavy mesons into two 
pseudo-scalar mesons ($P_{1,2} = K,\pi$). 
In this work the electromagnetic interactions of hadrons are described by minimal 
coupling of the photon field to point-like pseudoscalar meson fields.  
The authors perform a complete $O(\alpha)$ calculation, which correctly captures 
the potentially large IR effects due to virtual and real photons. There is no attempt to perform 
a UV matching and to model the "direct emission" of real photons.  
This implies an overall  $O(\alpha/\pi)$ uncertainty in the final results and the impossibility to 
apply the results to emission of hard photons.

\begin{figure}[h]
\centering
\includegraphics[width=80mm]{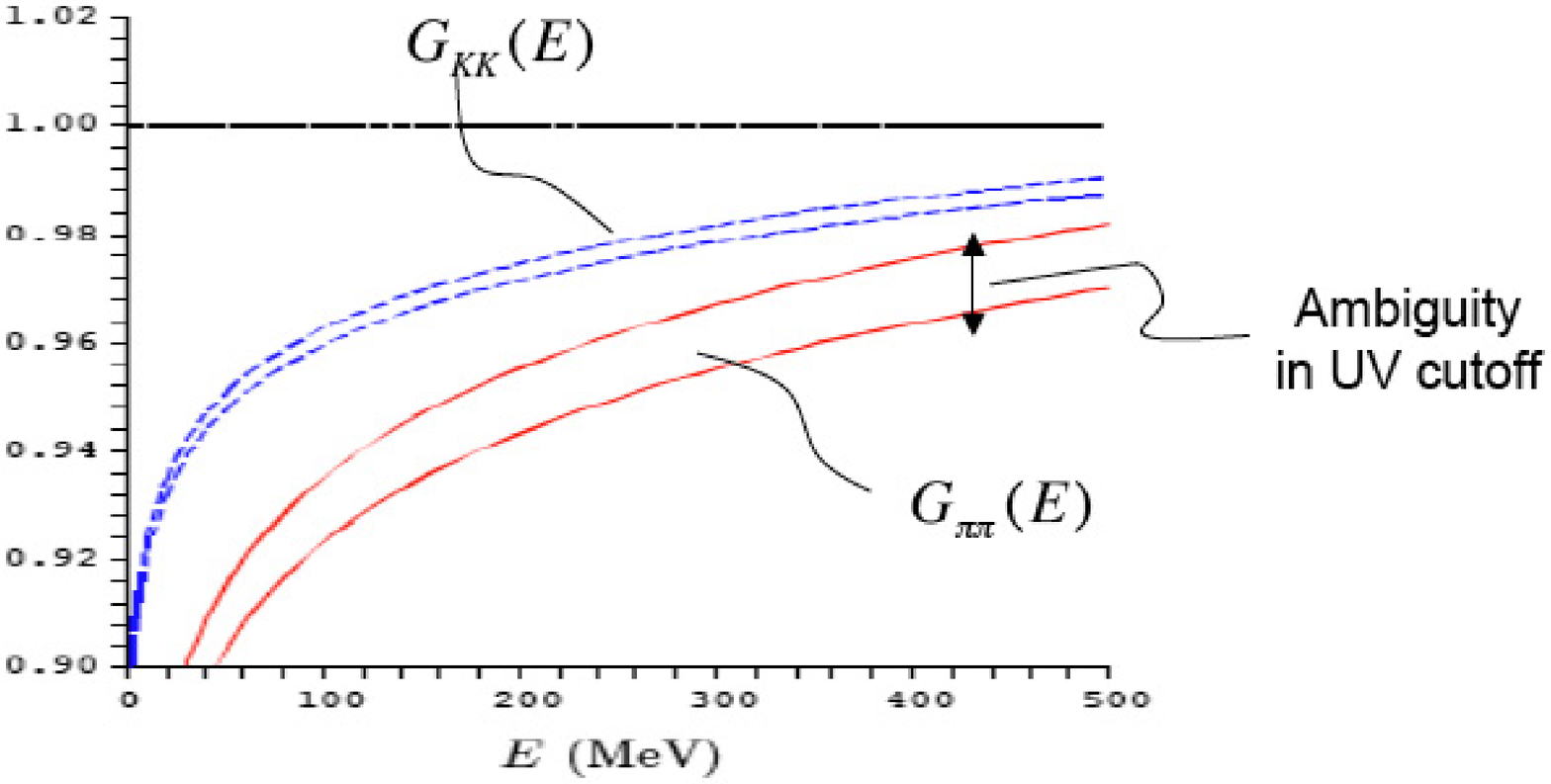}
\caption{
Examples of the~e.m.~correction factors for the photon-inclusive 
widths as a function of the cut on the maximal missing energy~\cite{Baracchini:2005wp}.
The full lines represent  $G_{\pi^+\pi^-}$ while the dashed lines 
represent $G_{K^+K^-}$
for the decays of $B$ mesons. 
In each case the upper curve is obtained by taking the renormalization scale 
$\mu = M_\rho$  while the lower curve for 
$\mu = M_\pi (M_K)$.
The difference is an estimate of the ambiguity entailed in the UV matching. 
} \label{fig:Gfact}
\end{figure}

Despite these limitations, the analysis of Ref.~\cite{Baracchini:2005wp}  has provided  
a definite improvement  over existing calculations and  MC implementations 
of radiative corrections for non-leptonic modes of heavy mesons. 
Moreover, it nicely  illustrates the need to define an IR-safe observable, 
i.e. the necessity to give a prescription for  the treatment of soft photons that 
always accompany charged decay products.  
The simplest definition of  IR-safe observable~\cite{Baracchini:2005wp} 
involves a sum over the parent and radiative modes, with an upper cutoff 
on the energy carried by photons in the CMS of the decaying particle:
\begin{equation}
\Gamma^{\rm incl} _{H \to P_1 P_2} (E^{\rm max}) = 
\sum_n \Gamma (H \to P_1 P_2 + n \gamma )  \Big|_{\sum E_\gamma < E^{\rm max}}~.
\end{equation}
With the above definition, the explicit calculation leads to 
\begin{equation}
\Gamma^{\rm incl} _{H \to P_1 P_2} (E^{\rm max})
= \Gamma^{0} _{H \to P_1 P_2}  \, 
 G_{12} (E^{\rm max})
\end{equation}
where $\Gamma^{0} _{H \to P_1 P_2}$  represents  the {\em unobservable} 
purely weak decay rate and $G_{12} (E^{\rm max})$ is the IR correction factor given 
to $O(\alpha)$ by:
\begin{equation}
G_{12} (E^{\rm max}) = 1 + b  \, \log \left( \frac{M_H}{2 E^{\rm max}} \right) 
+ c + O \left( 
\frac{E^{\rm max}}{M_H} 
\right)
\end{equation}
where $b$ is the factor of Eq.~\ref{eq:b} for the relevant decay mode and $c$ is 
a constant of  $O(\alpha)$ fixed by the complete calculation of Ref.~\cite{Baracchini:2005wp} .  
In Fig.~\ref{fig:Gfact} I report  the results for $G_{\pi \pi}$ and $G_{KK}$ in $B$ 
decays~\cite{Baracchini:2005wp}. The plot illustrates the importance to declare 
how the measured inclusive rate is defined: different definitions 
imply that one has to use  different correction factors $G_{12} (E^{\rm max})$ 
to extract $\Gamma^0$ (and ultimately learn about the underlying weak and strong dynamics). 
For example,  the correction factor is as large as $5\%$ in $B ^0 \to \pi^+ \pi^-$  
when $E^{\rm max} = 250$ MeV.

\subsection{Factorization in  $B \to \pi \ell  \bar{\nu}_\ell  \gamma$} 

Among radiative decays of heavy mesons, 
the  mode $B\to \gamma \pi \ell \bar\nu_\ell$  offers one of the simplest  
testing-grounds of EFT tools.  
The hadronic dynamics of the radiative semileptonic decays is
complicated by the presence of many energy scales. However, in certain
kinematic regions a model independent treatment is possible.  In the
limit of a soft photon, Low's theorem \cite{Low} fixes the terms
of order $O(q^{-1})$ and $O(q^0)$ in the expansion of the decay
amplitude in powers of the photon momentum $q$. 
Higher order terms in $q$ could be calculated  with the help of the heavy hadron
chiral perturbation theory \cite{wise,BuDo,TM,review}, as long as the
pion and photon are sufficiently soft in the B rest-frame.  This is the case in a limited
corner of the phase space, corresponding to a large dilepton invariant
mass $W^2=(p_{e}+p_\nu)^2 \sim m_b^2$.
On the other hand,  at low $W^2$ 
the  decay $B\to \gamma \pi \ell \bar\nu_\ell$ 
produces in general either an energetic pion or an energetic photon (see Fig.~1). 
The non-trivial dynamics is encoded in the  correlator of the weak semileptonic current with the 
electromagnetic current   ($j_\mu^{\rm em}=\sum_q e_q \bar q\gamma_\mu q$)
\begin{eqnarray}\label{corr}
T_{\mu\nu}(q) =
i\int d^4 x e^{iq.x} T \{(\bar q\gamma_\nu P_L b)(0)\,, j_\mu^{\rm e.m.}(x) \} . 
\end{eqnarray}
Using soft-collinear effective theory (SCET)~\cite{scet0,bfps,scet2,scet3}-\cite{scet3.5,scet4} 
methods,  it is possible to prove  a factorization relation for the decay amplitude
{\em in the kinematic region with an energetic photon and a soft pion}~\cite{Cirigliano:2005ms}.
The hadronic matrix element of the correlator defined in Eq.~\ref{corr} can then be written 
as 
\begin{eqnarray}\label{corr2}
& \langle \pi (p') | T_{\mu\nu}(q) | \bar{B} (p) \rangle =
C^{a}  (E_\gamma)  \times  \nonumber \\
& \int d\omega \, J (\omega) \,  \times  \langle \pi (p') | O^a_{\mu\nu}(\omega) | \bar{B} (p) \rangle
  + O(\Lambda/Q) ~ , &  
\label{1}
\end{eqnarray}
with $\Lambda \sim \Lambda_{QCD}$ and 
$Q \sim \{E_\gamma , m_b \}$. Here $C^a$ and  $J$ denote  Wilson coefficients 
encoding the corrections coming  respectively from  hard and hard-collinear 
degrees of freedom and are therefore calculable in perturbation theory.
$O^a_{\mu \nu}$ denote  bi-local operators expressed in terms of 
$b$ and light-quark fields as well as  soft Wilson lines. 
The above factorization relation holds for  any process $B\to
\gamma M\ell \nu$ ($M = \rho, \omega, \cdots$ a light soft meson) 
with the appropriate soft function $S(B\to M)$ replacing 
$\langle \pi (p') | O^a_{\mu\nu}(\omega) | \bar{B} (p) \rangle$. 
The key point is that  when $M=\pi$  we can use heavy hadron chiral perturbation
theory~\cite{wise,BuDo,TM,review} to relate 
the soft matrix elements $\langle \pi (p') | O^a_{\mu\nu}(\omega) | \bar{B} (p) \rangle$ 
to $\langle 0 | O^a_{\mu\nu}(\omega) | \bar{B} (p) \rangle$, that in turn can be 
expressed in terms of the B meson light-cone wave function    
$\phi_+^B(k_+)$~\cite{GPchiral}.  
The convolution of  $\phi_+^B(k_+)$  with the jet function $J$ in Eq.~(\ref{1})   
is the main non-perturbative parameter characterizing 
$B\to \gamma \pi \ell \bar\nu_\ell$
to leading order in $\Lambda/Q$.  
\begin{figure}[b!]
\begin{tabular}{cc}
{\includegraphics[height=5cm]{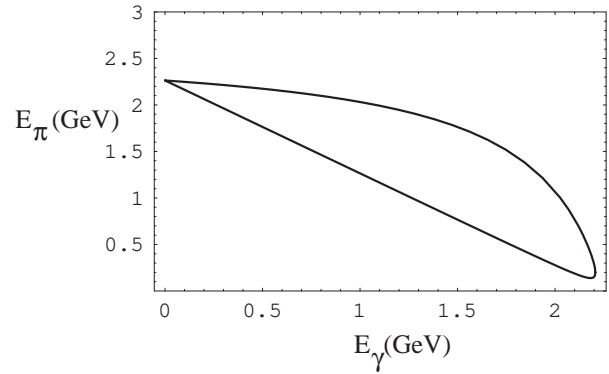}}
\end{tabular}
\caption{\label{fig1} 
The Dalitz plot for $B\to \gamma\pi \ell\bar\nu$ at $W^2=4$ GeV$^2$.
Our computation of the form factors is applicable in the lower right corner of
the Dalitz plot, with an energetic photon and a soft pion.
}
\end{figure}

In Ref.~\cite{Cirigliano:2005ms}  we analyzed 
the above  factorization formula, with a twofold intent:
(i) give predictions for quantities such as photon spectrum 
and the integrated rate with appropriate cuts;
(ii) compare our effective-theory based results with simplified approaches to 
$\bar{B} \to \pi e \bar{\nu} \gamma$, which are usually implemented 
in the Monte Carlo (MC) simulations used in the experimental analysis. 
In the numerical study we focused on  the neutral B decay
$\bar{B}^0 \to \pi^+ e^- \bar{\nu}_e \gamma$.

\noindent \underline{Factorization results}\\
The theoretical tools used (SCET and HHChPT) force  us to provide
 results for the photon energy spectrum, the 
distribution in the electron-photon angle, and integrated 
rate with various kinematic cuts:
\begin{enumerate}
\item An upper cut on the pion energy in the B rest frame 
$E_\pi < E_\pi^{\rm max}$,  required by the
applicability of chiral perturbation theory.  
\item  A lower cut on the
charged lepton-photon angle in the rest frame of the $B$ meson,  
$\theta_{e \gamma} > \theta^{\rm min}_{e \gamma}$, required   
to regulate the collinear singularity introduced
by photon radiation off the charged lepton. 
\item  Finally,  
to ensure the validity of the factorization theorem,
we require the photon to be sufficiently energetic (in the $B$ rest frame)  
$E_\gamma^{\rm min} \gg \Lambda_{QCD}$.
We choose $E_\gamma^{\rm min} = 1$ GeV.
\end{enumerate} 

Working to leading order in $\alpha_s$ and in the chiral expansion,
the non-perturbative  quantities needed are:   the first inverse moment of the 
light-cone B wave function  $\phi_+^B(k_+)$ 
\begin{equation}
\lambda_B^{-1}= \int dk_+  \  \frac{\phi_B^+(k_+,\mu)}{k_+} \ , 
\end{equation} 
the $B$ meson decay constant $f_B$, and the  HHChPT  $B-B^*-\pi$ coupling $g$. 
The numerical ranges used for 
the various input parameters \cite{parameters,parameters1,parameters2} 
are summarized in Table~\ref{table}. 

\begin{figure*}[!t]
\centering
\begin{picture}(300,170)  
%\put(-10,55){\makebox(50,50){\epsfig{figure=spectra2.eps,height=5cm}}}
\put(-10,55){\makebox(50,50){
\includegraphics[height=5cm]{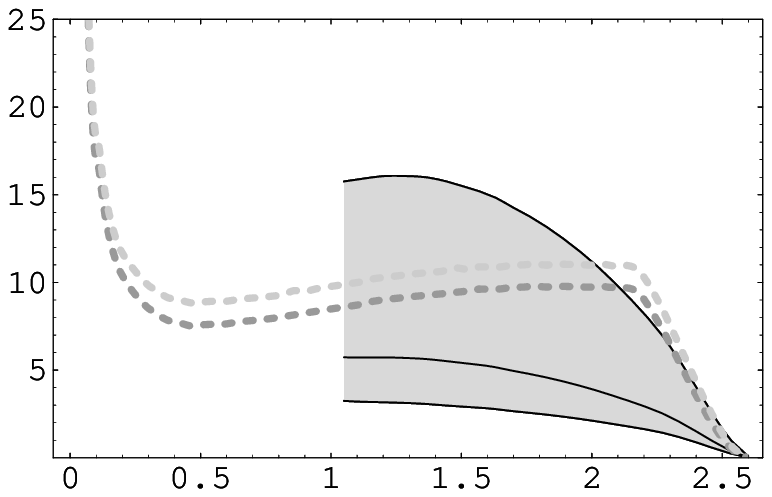}
}}
\put(263,52){\makebox(50,50){
\includegraphics[height=5cm]{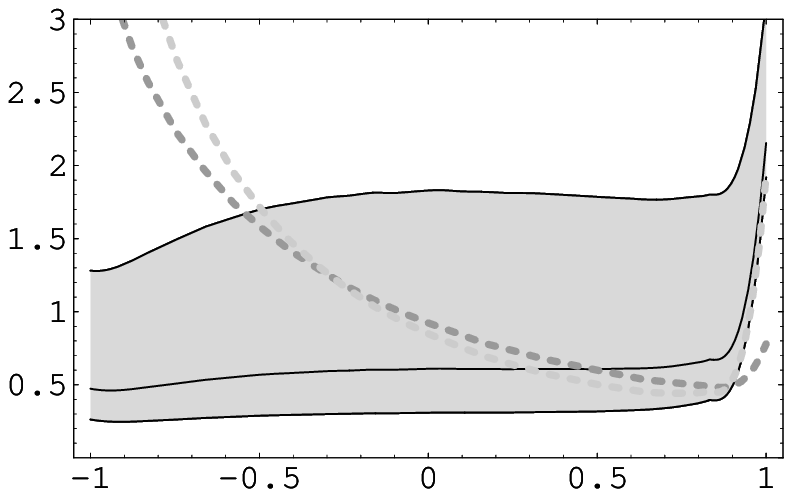}
}}
%\put(263,52){\makebox(50,50){\epsfig{figure=ang2.eps,height=5cm}}}
\put(-40,155){
\framebox[1.0\width][c]{
\small $ E_\pi < 0.5 \ {\rm GeV} \qquad  \theta_{e \gamma} > 5^o  $  
}}
\put(230,155){
\framebox[1.0\width][c]{
\small $ E_\pi < 0.5 \ {\rm GeV} \qquad  E_\gamma > 1 \ {\rm GeV}  $  
}}
\put(-115,85){
{\Large 
$\frac{d \bar{\Gamma}}{d E_\gamma}$
}
}
\put(128,85){
{\Large 
$\frac{d \bar{\Gamma}}{d (\cos \theta_{e \gamma})}$
}
}
\put(280,-5){
{\large 
$\cos \theta_{e \gamma}$ 
}
}
\put(-5,-5){
{\large 
$E_\gamma$ (GeV) 
}
}
\put(-20,72){
{\small
IB2
}
}
\put(-35,48){
{\small
IB1
}
}
\put(30,50){
{\small
FACT  
}
}
%%%%%%%%%%%%%%%%%%%%%%%%%%%%%%%%%%%%%%%%%%%%%%%%%%%%%%%%%%%%%
\put(220,45){
{\small
FACT
}
}
\put(210,90){
{\small
IB1
}
}
\put(245,100){
{\small
IB2
}
}
\end{picture}
\caption{\label{fig:spectra2} 
Left panel: Differential rate 
$d \bar{\Gamma}/d E_\gamma$, as a function of the photon energy in the
$B$-meson rest frame.  The normalization is given by $\bar{\Gamma}=
10^6 \, \Gamma_{B^0}^{-1} \times \Gamma (\bar{B} \to \pi e \bar{\nu}
\gamma)$.  Right panel: Differential radiative rate 
$d \bar{\Gamma}/d ( \cos \theta_{e \gamma} )$, as a function of the
cosine of electron-photon angle in the  $B$-meson rest frame.
The shaded bands and the superimposed continuous lines
represent the uncertainty in the factorization
result (FACT) induced by the parameter $\lambda_B$, assumed to range
between 200 (upper line) and 500 (lower line) MeV, 
with all other parameters equal to the central values of Table~\ref{table}. 
The dark(light)-gray dashed line corresponds
to the IB1 (IB2) results (see text for their definition).  
}
\end{figure*}

The phase space integration with the cuts described above was
performed using the Monte Carlo event generator
RAMBO~\cite{RAMBO}. The factorization results for the photon energy
spectrum and the distribution in $\cos \theta_{e \gamma}$ are shown in
Fig.~\ref{fig:spectra2} (solid lines), using
the central and extreme values for the input parameters as reported in
Table~\ref{table}. The shaded bands reflect the uncertainty in our prediction,
which is due mostly to the variation of $\lambda_B$.
For $E_\gamma^{\rm min}=1 \ {\rm GeV}$, $E_\pi^{\rm max}=0.5 \ {\rm GeV}$, 
and $\theta_{e \gamma}^{\rm min}=5^o$, 
the integrated radiative branching fraction predicted by our factorization 
formula is (up to small perturbative corrections in $\alpha_s(m_b)$ and 
$\alpha_s(\sqrt{m_b \Lambda})$)
\begin{eqnarray}
{\rm Br}_{\bar{B} \to \pi e \bar{\nu} \gamma}^{\rm cut} ({\rm fact}) 
&=& 
 \Big(1.2 \  \pm 0.2 (g)  \ ^{+ 2.2} _{-0.6} (\lambda_B)  \Big) 
\times 10^{-6} 
\nonumber \\
&\times & \left(\frac{|V_{ub}|}{0.004}\right)^2  \times
\left(\frac{f_B}{200 \ {\rm MeV}}\right)^2  \ . 
\label{eq:brfact}
\end{eqnarray}
Apart from  the overall quadratic dependence on $V_{ub}$ and $f_B$, 
the main uncertainty of this prediction comes from the  poorly known 
non-perturbative parameter $\lambda_B$. 
It is worth mentioning  that radiative rate, photon spectrum and
angular distribution could be used in the future to obtain
experimental constraints on $\lambda_B$, given their strong
sensitivity to it.  This requires, however, a reliable estimate of the
impact of chiral corrections, $O(\alpha_s)$ corrections, and the
uncertainty in other input parameters.

\begin{table}%[H] add [H] placement to break table across pages
\caption{\label{table} Input parameters used in the numerical 
computation~\cite{parameters,parameters1,parameters2}.}
%\begin{ruledtabular}
\begin{tabular}{|cc|cc|}
\hline
$f_B$ & $(200\pm 30)$ MeV  & $f_\pi$ & $131$ MeV \\
$\lambda_B$  & $(350 \pm 150)$ MeV  &  $g$   & $0.5 \pm 0.1$  \\
$|V_{ub}|$ & $0.004$ &  &   \\
\hline
\end{tabular}
%\end{ruledtabular}
\end{table}

\begin{table}%[H] add [H] placement to break table across pages
\caption{\label{table2} Branching fractions of the 
cut $\bar{B} \to \pi e \bar{\nu} \gamma$ decay 
% in units of $10^{-6}\times |V_{ub}/0.004|^2$, 
corresponding to the factorization amplitude and the IB1, IB2 cases 
defined in the text, 
for central values of the input parameters reported in Table~\ref{table}. 
%%%%%%%%%%%%%%%%%%%%%%
%The last column shows the ratio $R$
%of the radiative to nonradiative branching fractions (defined in 
%Eq.~(\ref{Rdef})).
%%%%%%%%%%%%%%%%%%%%%%%
}
%\begin{ruledtabular}
\begin{tabular}{|c|ccc|}
% &  &  &  \\
%\hline
%cuts  & Br(IB1)  & Br(FACT) & $R$  \\
\hline
cuts  & $10^6$  Br(fact)  & $10^6$ Br(IB1) & $10^6$ Br(IB2)  \\
\hline
$E_\gamma > 1$ GeV &       &        &  \\
$E_\pi < 0.5$ GeV & $1.2$ & $2.4$ & $2.8$ \\
$\theta_{e\gamma}>5^\circ$ &       &        &  \\
\hline
\end{tabular}
%\end{ruledtabular}
\end{table}

\noindent \underline{Comparison with simplified approaches} \\
In the Monte Carlo simulations used in most experimental analyses of B
decays, simplified versions of the radiative hadronic matrix elements
are implemented. These rely essentially on Low's theorem~\cite{Low}  
and are valid only in the limit of soft photons
($E_\gamma \ll \Lambda$).  
They are nonetheless used over the entire kinematic range, for 
lack of better results. 
Our interest here is to evaluate how these methods fare in the region
of soft pion and hard photon (certainly beyond their regime of
validity), where we can use the QCD-based factorization results as 
benchmark.  

We use two simplified versions of the $\bar{B} \to \pi e \bar{\nu}
\gamma$ matrix element, which we denote  IB1 and IB2. 
%They arise as special cases of  Low's amplitude: 
%
\begin{itemize}
\item[IB1:] This is the $O(q^{-1})$ component of Low's amplitude. 
The PHOTOS Monte Carlo~\cite{photos} implements this form of the
radiative amplitude.
\item[IB2:] This corresponds to a tree-level calculation with point-like 
$B$ and $\pi$,   and a constant transition form factor~\cite{Ginsberg}. 
\end{itemize}
The results for the photon energy spectrum and the distribution in
$\cos \theta_{e \gamma}$ are shown in Figs.~\ref{fig:spectra2} 
(IB1: dark-gray dashed lines, IB2: light-gray
dashed lines).  The integrated branching fractions are listed in
Table~\ref{table2} (together with the factorization prediction) for
central values of the input parameters given in Table~\ref{table}.
The IB1 and IB2 branching fractions scale very simply with the input
parameters, being proportional to $(|V_{ub}| \, g \, f _B)^2$.
As can be seen from the comparison of differential distributions and
BRs,  the two cases IB1 and IB2 lead to almost identical predictions,
which are, however,  quite different
from the factorization ones.  For central values of
the input parameters reported in Table~\ref{table} the IB1 and IB2
predict a radiative BR roughly a factor of two 
larger than the factorization
result. Moreover, the shape of the distributions in $E_\gamma$ and
$\cos \theta_{e \gamma}$ are quite different. 
We hope that our results can help construct more reliable MCs for
the simulation of $\bar{B} \to \pi e \bar{\nu}\gamma$ decays, 
or at least address the systematic uncertainty to be assigned to the 
present MC results.

\section{Conclusions}

In this talk I have reviewed recent developments in the calculation of 
radiative corrections to $K$ and $B$ decays, arguing that  
control of radiative effects is mandatory  in order to make 
meaningful phenomenology at the few $\%$ level. 

The situation in $K$ decays is satisfactory, because the chiral
expansion provides a robust tool to calculate radiative corrections
and estimate the residual uncertainties.

In $B$ and $D$ decays there is no "universal" effective theory that
can be invoked.  Rigorous treatments are possible only in limited 
regions of phase space.  For those modes in which experimental precision
will reach the few $\%$ level in the near future, the minimal goal is
to perform complete $O(\alpha)$ calculation in the approximation of
point-like hadrons. This should at least capture the leading IR
effects, that are known to be large in some
modes~\cite{Baracchini:2005wp}.  Whenever possible, it is desirable to
perform calculations within the appropriate hadronic effective theory.
This will help assessing the uncertainty entailed in the point-like
approximation and in the Monte Carlo simulations presently used in
experimental analyses.  As an example of EFT calculation, I have
discussed the process $\bar{B} \to \pi e \bar{\nu}\gamma$ in the
regime of soft pion and hard photon~\cite{Cirigliano:2005ms}.

\bigskip % extra skip insertedi
\begin{acknowledgments}
I warmly thank the co-authors of Refs.~\cite{k2pi}, \cite{radcorr1}, \cite{radcorr1bis}, 
\cite{blucher-marciano}, and \cite{Cirigliano:2005ms}  
for stimulating collaborations. 
\end{acknowledgments}

\bigskip % extra skip inserted

\begin{thebibliography}{99} % Use for 10-99 references



\bibitem{RadCorr}
See website of {\em Workshop on Radiative Corrections in B, D, and K 
Meson Decays}, UCSD (La Jolla, CA), March 14 2005, 
www.slac.stanford.edu/BFROOT/www/Public/ \\
Organization/2005/workshops/radcorr2005/\\ index.html. 

\bibitem{CGatti}
C.~Gatti,
  %``Monte Carlo simulation for radiative kaon decays,''
  arXiv:hep-ph/0507280.
  %%CITATION = HEP-PH 0507280;%%

\bibitem{SDcorr1}
% \bibitem{Sirlin:1981ie}
  A.~Sirlin,
  %``Large M (W), M (Z) Behavior Of The O (Alpha) Corrections To Semileptonic
  %Processes Mediated By W,''
  Nucl.\ Phys.\ B {\bf 196}, 83 (1982). 
  %%CITATION = NUPHA,B196,83;%%

\bibitem{SDcorr2} 
%\cite{Lusignoli:1988fz}
%\bibitem{Lusignoli:1988fz}
  M.~Lusignoli,
  %``Electromagnetic Corrections To The Effective Hamiltonian For Strangeness
  %Changing Decays And Epsilon-Prime / Epsilon,''
  Nucl.\ Phys.\ B {\bf 325}, 33 (1989); 
  %%CITATION = NUPHA,B325,33;%%
%
%\cite{Buras:1992tc}
%\bibitem{Buras:1992tc}
  A.~J.~Buras, M.~Jamin, M.~E.~Lautenbacher and P.~H.~Weisz,
  %``Two loop anomalous dimension matrix for Delta S = 1 weak nonleptonic
  %decays. 1. O(alpha-s**2),''
  Nucl.\ Phys.\ B {\bf 400}, 37 (1993)
  [arXiv:hep-ph/9211304]; 
  %%CITATION = HEP-PH 9211304;%%
%
%\cite{Ciuchini:1993vr}
%\bibitem{Ciuchini:1993vr}
  M.~Ciuchini, E.~Franco, G.~Martinelli and L.~Reina,
  %``The Delta S = 1 effective Hamiltonian including next-to-leading order QCD
  %and QED corrections,''
  Nucl.\ Phys.\ B {\bf 415}, 403 (1994)
  [arXiv:hep-ph/9304257].
  %%CITATION = HEP-PH 9304257;%%


%\cite{Weinberg:1965nx}\cite{Yennie:1961ad}
\bibitem{Weinberg:1965nx}
  S.~Weinberg,
  %``Infrared Photons And Gravitons,''
  Phys.\ Rev.\  {\bf 140} (1965) B516.
  %%CITATION = PHRVA,140,B516;%%


%\cite{Yennie:1961ad}
\bibitem{Yennie:1961ad}
  D.~R.~Yennie, S.~C.~Frautschi and H.~Suura,
  %``The Infrared Divergence Phenomena And High-Energy Processes,''
  Annals Phys.\  {\bf 13} (1961) 379.
  %%CITATION = APNYA,13,379;%%



%%%%%%%%%%%%% ChPT with virtual photons   %%%%%%%%%%%%

%\cite{Knecht:1999ag}\cite{Ecker:2000zr}
\bibitem{Knecht:1999ag}
  M.~Knecht, H.~Neufeld, H.~Rupertsberger and P.~Talavera,
  %``Chiral perturbation theory with virtual photons and leptons,''
  Eur.\ Phys.\ J.\ C {\bf 12}, 469 (2000)
  [arXiv:hep-ph/9909284].
  %%CITATION = HEP-PH 9909284;%%


%\cite{Ecker:2000zr}
\bibitem{Ecker:2000zr}
  G.~Ecker, G.~Isidori, G.~Muller, H.~Neufeld and A.~Pich,
  %``Electromagnetism in nonleptonic weak interactions,''
  Nucl.\ Phys.\ B {\bf 591}, 419 (2000)
  [arXiv:hep-ph/0006172].
  %%CITATION = HEP-PH 0006172;%%



%%%%%%%%%% K -> 2 pi  %%%%%%%%%%%%%%

\bibitem{k2pi}
%\cite{Cirigliano:1999ie}
%\bibitem{Cirigliano:1999ie}
  V.~Cirigliano, J.~F.~Donoghue and E.~Golowich,
  %``Electromagnetic corrections to K $\to$ pi pi. I: Chiral perturbation
  %theory,''
  Phys.\ Rev.\ D {\bf 61}, 093001 (2000)
  [Erratum-ibid.\ D {\bf 63}, 059903 (2001)]
  [arXiv:hep-ph/9907341];
  %%CITATION = HEP-PH 9907341;%%
%
%
%\cite{Cirigliano:2000zw}
%\bibitem{Cirigliano:2000zw}
%  V.~Cirigliano, J.~F.~Donoghue and E.~Golowich,
  %``K $\to$ pi pi phenomenology in the presence of electromagnetism,''
  Eur.\ Phys.\ J.\ C {\bf 18}, 83 (2000)
  [arXiv:hep-ph/0008290]. \\
  %%CITATION = HEP-PH 0008290;%%
%
%
%
%\cite{Cirigliano:2003nn}
%\bibitem{Cirigliano:2003nn}
  V.~Cirigliano, A.~Pich, G.~Ecker and H.~Neufeld,
  %``Isospin violation in epsilon',''
  Phys.\ Rev.\ Lett.\  {\bf 91}, 162001 (2003)
  [arXiv:hep-ph/0307030].
  %%CITATION = HEP-PH 0307030;%%
%
%\cite{Cirigliano:2003gt}
%\bibitem{Cirigliano:2003gt}
%  V.~Cirigliano, G.~Ecker, H.~Neufeld and A.~Pich,
  %``Isospin breaking in K $\to$ pi pi decays,''
  Eur.\ Phys.\ J.\ C {\bf 33}, 369 (2004)
  [arXiv:hep-ph/0310351].
  %%CITATION = HEP-PH 0310351;%%


%%%%%%%%%% K -> 3 pi  %%%%%%%%%%%%%%

\bibitem{k3pi}
%\cite{Bijnens:2004vz}
%\bibitem{Bijnens:2004vz}
  J.~Bijnens and F.~Borg,
  %``Isospin breaking in K $\to$ 3 pi decays. II: Radiative corrections,''
  Eur.\ Phys.\ J.\ C {\bf 39}, 347 (2005)
  [arXiv:hep-ph/0410333];
  %%CITATION = HEP-PH 0410333;%%
%
%
%\cite{Bijnens:2004ai}
%\bibitem{Bijnens:2004ai}
% J.~Bijnens and F.~Borg,
  %``Isospin breaking in K $\to$ 3pi decays. III: Bremsstrahlung and fit to
  %experiment,''
  Eur.\ Phys.\ J.\ C {\bf 40}, 383 (2005)
  [arXiv:hep-ph/0501163].
  %%CITATION = HEP-PH 0501163;%%

%%%%%%%%%%%%%     Kl4       %%%%%%%%%%%%

\bibitem{kl4} 
%\cite{Cuplov:2003bj}
%\bibitem{Cuplov:2003bj}
  V.~Cuplov and A.~Nehme,
  %``Isospin breaking in K(l4) decays of the charged kaon,''
  arXiv:hep-ph/0311274; 
  %%CITATION = HEP-PH 0311274;%%
%
%\cite{Nehme:2003bz}
%\bibitem{Nehme:2003bz}
  A.~Nehme,
  %``Isospin breaking in K(l4) decays of the neutral kaon,''
  Nucl.\ Phys.\ B {\bf 682}, 289 (2004)
  [arXiv:hep-ph/0311113].
  %%CITATION = HEP-PH 0311113;%%


%%%%%%%%%%%%%%%%%%%%%%%%%%%%%%%%%%%%%%%



\bibitem{Weinberg:1978kz}
%\cite{Weinberg:1978kz}
% \bibitem{Weinberg:1978kz}
S.~Weinberg,
%``Phenomenological Lagrangians,''
Physica A {\bf 96} (1979) 327.
%%CITATION = PHYSA,A96,327;%%

\bibitem{Gasser:1983yg}
%\cite{Gasser:1983yg}
%\bibitem{Gasser:1983yg}
J.~Gasser and H.~Leutwyler,
%``Chiral Perturbation Theory To One Loop,''
Ann. Phys.\  {\bf 158} (1984) 142.
%%CITATION = APNYA,158,142;%%

\bibitem{Gasser:1984gg}
J.~Gasser and H.~Leutwyler,
%``Chiral Perturbation Theory: Expansions In The Mass Of The Strange Quark,''
Nucl.\ Phys.\ B {\bf 250} (1985) 465.
%%CITATION = NUPHA,B250,465;%%

%\cite{Gasser:1984ux}
\bibitem{Gasser:1984ux}
J.~Gasser and H.~Leutwyler,
%``Low-Energy Expansion Of Meson Form-Factors,''
Nucl.\ Phys.\ B {\bf 250} (1985) 517.
%%CITATION = NUPHA,B250,517;%%

\bibitem{radcorr1}
V.~Cirigliano, M.~Knecht, H.~Neufeld, H.~Rupertsberger and P.~Talavera,
%``Radiative corrections to K(l3) decays,''
Eur.\ Phys.\ J.\ C {\bf 23} (2002) 121
[hep-ph/0110153].

\bibitem{radcorr1bis}
V.~Cirigliano, H.~Neufeld and H.~Pichl,
%``K(e3) decays and CKM unitarity,''
Eur.\ Phys.\ J.\ C {\bf 35} (2004) 53 [hep-ph/0401173].
%%CITATION = HEP-PH 0401173;%%


\bibitem{moussallam}
%\cite{Moussallam:1997xx}
%\bibitem{Moussallam:1997xx}
 B.~Moussallam,
%``A sum rule approach to the violation of Dashen's theorem,''
Nucl.\ Phys.\ B {\bf 504}, 381 (1997)
[hep-ph/9701400]; 
%%CITATION = HEP-PH 9701400;%%
%\cite{Ananthanarayan:2004qk}
%\bibitem{Ananthanarayan:2004qk}
  B.~Ananthanarayan and B.~Moussallam,
  %``Four-point correlator constraints on electromagnetic
  %  chiral parameters  and
  % resonance effective Lagrangians,''
  JHEP {\bf 0406}, 047 (2004)
  [arXiv:hep-ph/0405206].
  %%CITATION = HEP-PH 0405206;%%

\bibitem{descotes}
%\cite{Descotes-Genon:2005pw}
%\bibitem{Descotes-Genon:2005pw}
  S.~Descotes-Genon and B.~Moussallam,
  %``Radiative corrections in weak semi-leptonic processes at low energy: A
  %two-step matching determination,''
  Eur.\ Phys.\ J.\ C {\bf 42}, 403 (2005)
  [arXiv:hep-ph/0505077].
  %%CITATION = HEP-PH 0505077;%%


\bibitem{kl3rad} 
J.~Bijnens, G.~Ecker and J.~Gasser,
%``Radiative semileptonic kaon decays,''
Nucl.\ Phys.\ B {\bf 396}, 81 (1993)
[hep-ph/9209261]; \\
%%CITATION = HEP-PH 9209261;%%
%\cite{Gasser:2004ds}
%\bibitem{Gasser:2004ds}
J.~Gasser, B.~Kubis, N.~Paver and M.~Verbeni,
%``Radiative K(e3) decays revisited,''
Eur.\ Phys.\ J.\ C {\bf 40}, 205 (2005)
[hep-ph/0412130].
%%CITATION = HEP-PH 0412130;%%

\bibitem{radcorr2}
T.~C.~Andre,
%``Radiative corrections to K0(l3) decays,''
hep-ph/0406006. 
%%CITATION = HEP-PH 0406006;%%

\bibitem{radcorr3}
%\cite{Bytev:2002nx}
%\bibitem{Bytev:2002nx}
V.~Bytev, E.~Kuraev, A.~Baratt and J.~Thompson,
%``Radiative corrections to the K+-(e3) decay revised,''
Eur.\ Phys.\ J.\ C {\bf 27} (2003) 57
[Erratum-ibid.\ C {\bf 34} (2004) 523]
[hep-ph/0210049].
%%CITATION = HEP-PH 0210049;%%

%\cite{Blucher:2005dc}\cite{Cirigliano:2005ms}
%\bibitem{Blucher:2005dc}
 \bibitem{blucher-marciano}
  E.~Blucher {\it et al.},
  %``Status of the Cabibbo angle (CKM2005 - WG 1),''
  arXiv:hep-ph/0512039.
  %%CITATION = HEP-PH 0512039;%%

%%%%%%%%%%%%%%%%%%%%%%%%%%%%%%%%%%%%%%%%%%%%%%%%

%\cite{Baracchini:2005wp}
\bibitem{Baracchini:2005wp}
  E.~Baracchini and G.~Isidori,
  %``Electromagnetic corrections to non-leptonic two-body B and D decays,''
  Phys.\ Lett.\ B {\bf 633}, 309 (2006)
  [arXiv:hep-ph/0508071].
  %%CITATION = HEP-PH 0508071;%%

%%%%%%%%%%%%%%%%%%%%%%%%%%%%%%%%%%%%%%%%%%%%%%%%

%\cite{Cirigliano:2005ms}
\bibitem{Cirigliano:2005ms}
  V.~Cirigliano and D.~Pirjol,
  %``Factorization in exclusive semileptonic radiative B decays,''
  Phys.\ Rev.\ D {\bf 72}, 094021 (2005)
  [arXiv:hep-ph/0508095].
  %%CITATION = HEP-PH 0508095;%%


\bibitem{Low}
F.~E.~Low,
  %``Bremsstrahlung Of Very Low-Energy Quanta In Elementary Particle
  %Collisions,''
  Phys.\ Rev.\  {\bf 110}, 974 (1958).
  %%CITATION = PHRVA,110,974;%%


\bibitem{wise}
M.~B.~Wise,
%``Chiral perturbation theory for hadrons containing a heavy quark,''
Phys.\ Rev.\ D {\bf 45}, 2188 (1992).
%%CITATION = PHRVA,D45,2188;%%

\bibitem{BuDo}
G.~Burdman and J.~F.~Donoghue,
%``Union of chiral and heavy quark symmetries,''
Phys.\ Lett.\ B {\bf 280}, 287 (1992).
%%CITATION = PHLTA,B280,287;%%

\bibitem{TM}
T.~M.~Yan {\em et al.},
%H.~Y.~Cheng, C.~Y.~Cheung, G.~L.~Lin, Y.~C.~Lin and H.~L.~Yu,
%``Heavy quark symmetry and chiral dynamics,''
Phys.\ Rev.\ D {\bf 46}, 1148 (1992)
[Erratum-ibid.\ D {\bf 55}, 5851 (1997)].
%%CITATION = PHRVA,D46,1148;%%

\bibitem{review}
M.~B.~Wise,
%``Combining chiral and heavy quark symmetry,''
arXiv:hep-ph/9306277.
%%CITATION = HEP-PH 9306277;%%

\bibitem{scet0}
C.~W.~Bauer, S.~Fleming and M.~E.~Luke,
%``Summing Sudakov logarithms in B $\to$ X/s gamma in effective field
%theory,''
Phys.\ Rev.\ D {\bf 63}, 014006 (2001).
%[arXiv:hep-ph/0005275].
%%CITATION = HEP-PH 0005275;%%


\bibitem{bfps}
%\bibitem{Bauer:2000yr}
C.~W.~Bauer, S.~Fleming, D.~Pirjol and I.~W.~Stewart,
%``An effective field theory for collinear and soft gluons: Heavy to light 
% decays,''
Phys.\ Rev.\ D {\bf 63}, 114020 (2001).
%[hep-ph/0011336];
%%CITATION = HEP-PH 0011336;%%

%\cite{Bauer:2001ct}
\bibitem{scet2}
%\bibitem{Bauer:2001ct}
C.~W.~Bauer and I.~W.~Stewart,
%``Invariant operators in collinear effective theory,''
Phys.\ Lett.\ B {\bf 516}, 134 (2001)
%[hep-ph/0107001];
%%CITATION = HEP-PH 0107001;%%

%\cite{Bauer:2001yt}
\bibitem{scet3}
C.~W.~Bauer, D.~Pirjol and I.~W.~Stewart,
%``Soft-collinear factorization in effective field theory,''
Phys.\ Rev.\ D {\bf 65}, 054022 (2002).
%[hep-ph/0109045].
%%CITATION = HEP-PH 0109045;%%

\bibitem{scet3.5}
J.~Chay and C.~Kim,
  %``Collinear effective theory at subleading order and its application to
  %heavy-light currents,''
  Phys.\ Rev.\ D {\bf 65}, 114016 (2002).
%[arXiv:hep-ph/0201197].
  %%CITATION = HEP-PH 0201197;%%

\bibitem{scet4}
M.~Beneke, A.~P.~Chapovsky, M.~Diehl and T.~Feldmann,
  %``Soft-collinear effective theory and heavy-to-light currents beyond  leading
  %power,''
  Nucl.\ Phys.\ B {\bf 643}, 431 (2002).
%  [arXiv:hep-ph/0206152].
  %%CITATION = HEP-PH 0206152;%%


\bibitem{GPchiral}
B.~Grinstein and D.~Pirjol,
%``Chiral symmetry and exclusive B decays in the SCET,''
Phys.\ Lett.\ B {\bf 615}, 213 (2005).
%[arXiv:hep-ph/0501237].
%%CITATION = HEP-PH 0501237;%%


\bibitem{parameters}
V.~M.~Braun, D.~Y.~Ivanov and G.~P.~Korchemsky,
  %``The B-meson distribution amplitude in QCD,''
  Phys.\ Rev.\ D {\bf 69}, 034014 (2004);
%  [arXiv:hep-ph/0309330];
  %%CITATION = HEP-PH 0309330;%%
A.~Khodjamirian, T.~Mannel and N.~Offen,
  %``B-meson distribution amplitude from the B $\to$ pi form factor,''
  Phys.\ Lett.\ B {\bf 620}, 52 (2005).
%  [arXiv:hep-ph/0504091].
  %%CITATION = HEP-PH 0504091;%%



\bibitem{parameters1}
C.~W.~Bernard,
  %``Heavy quark physics on the lattice,''
  Nucl.\ Phys.\ Proc.\ Suppl.\  {\bf 94}, 159 (2001)
  [arXiv:hep-lat/0011064].
  %%CITATION = HEP-LAT 0011064;%%

\bibitem{parameters2}
M.~C.~Arnesen, B.~Grinstein, I.~Z.~Rothstein and I.~W.~Stewart,
%``A precision model independent determination of $|$V(ub)$|$ from B $\to$ pi
%e nu,''
arXiv:hep-ph/0504209; 
%%CITATION = HEP-PH 0504209;%%
S.~Hashimoto and T.~Onogi,
%``Heavy quarks on the lattice,''
Ann.\ Rev.\ Nucl.\ Part.\ Sci.\  {\bf 54}, 451 (2004).
%[arXiv:hep-ph/0407221].
%%CITATION = HEP-PH 0407221;%%



\bibitem{RAMBO}
R.~Kleiss, W.~J.~Stirling and S.~D.~Ellis,
%``A New Monte Carlo Treatment Of Multiparticle Phase Space At
%High-Energies,''
Comput.\ Phys.\ Commun.\  {\bf 40}, 359 (1986).
%%CITATION = CPHCB,40,359;%%

\bibitem{photos} 
E.~Barberio and Z.~Was,
%``PHOTOS: A Universal Monte Carlo for QED radiative corrections. Version
%2.0,''
Comput.\ Phys.\ Commun.\  {\bf 79}, 291 (1994).
%%CITATION = CPHCB,79,291;%%


\bibitem{Ginsberg}
E.~S.~Ginsberg,
%``Radiative Corrections To K-Mu-3 Decays,''
Phys.\ Rev.\ D {\bf 1}, 229 (1970).
%%CITATION = PHRVA,D1,229;%%
%``Radiative Corrections To K-E-3-Neutral Decays And The Delta-I=1/2 Rule.
%(Erratum),''
Phys.\ Rev.\  {\bf 171}, 1675 (1968)
[Erratum-ibid.\  {\bf 174}, 2169 (1968)].
%%CITATION = PHRVA,171,1675;%%



\end{thebibliography}

%\begin{thebibliography}{9}   % Use for  1-9  references

\end{document}